# Testing Deep Learning Recommender Systems Models on Synthetic GAN-Generated Datasets

Jesús Bobadilla, Abraham Gutiérrez*

Universidad Politécnica de Madrid, Dpto. Sistemas Informáticos, Madrid (Spain)

*Corresponding author: abraham.gutierrez@upm.es

**Abstract**

GANRS is a published method for generating datasets for collaborative filtering recommender systems. The GANRS source code is available along with a representative set of generated datasets. We have tested the GANRS method by creating multiple synthetic datasets from three different real datasets taken as a source. Experiments include variations in the number of users in the synthetic datasets, as well as a different number of samples. We have also selected six state-of-the-art collaborative filtering deep learning models to test both their comparative performance and the GANRS method. The results show a consistent behavior of the generated datasets compared to the source ones; particularly, in the obtained values and trends of the precision and recall quality measures. The tested deep learning models have also performed as expected on all synthetic datasets, making it possible to compare the results with those obtained from the real source data. Future work is proposed, including different cold start scenarios, unbalanced data, and demographic fairness.

**Keywords:** Collaborative Filtering, Deep Learning, GANRS, Generated Datasets, Recommender Systems, Synthetic Datasets

## 1. Introduction

The personalization field in the Artificial Intelligence area is mainly focused on Recommender Systems (RS). Relevant RS are Netflix, TripAdvisor, Spotify, Google Music, TikTok, etc. RS are usually classified according to their filtering approaches, mainly: demographic [1], content-based [2], context-aware [3], social [4], collaborative (CF) [5] and their ensembles [6]. Demographic RS make recommendations based on demographic similarities (gender, age, zip code, etc.); content-based RS recommend items with similar content to the consumed ones (book abstracts, product images, etc.). Context-aware filtering usually uses geographic information, such as GPS coordinates. Social filtering relies on followed, followers, etc. CF uses datasets containing the ratings that each user has voted to each item. Ratings can be explicit votes or implicit interactions (clicks, music listened to, films watched, etc.). Of the existing filtering approaches, CF is the most relevant since it provides the most accurate results. The early approaches to CF used the K-Nearest Neighbors algorithm [7]; it is easy to understand and directly implements the concept of CF, but it is also a slow memory-based method, and its results are not accurate compared to modern model-based approaches. The Matrix Factorization (MF) model [8] creates compressed representations of the input data, called hidden factors, and then combines these latent space vectors using the dot product to obtain each user to item prediction. Probabilistic MF and its variations (NMF [9], BNMF, etc.) provide straightforward models that return accurate prediction and recommendations. Furthermore, once the MF model has been trained, it can make very fast predictions compared to the KNN method.

Currently, deep learning approaches dominate the RS research scenario. The simplest deep learning CF model is the Deep Matrix Factorization (DeepMF) [10], where iterative MF learning is replaced with two different neural embedding layers: one for code users and the other for code items. The embedding layers activation maps play the role of the MF hidden factors, where large, discrete, and sparse input vectors are converted to short, continuous, and dense latent space vectors. As in the MF case, the embedding vectors are combined using a dot layer. The variational design of the DeepMF model is called VDeepMF [11], where a Gaussian stochastic noise is introduced after the embedding layers to obtain more robust results. Neural Collaborative Filtering (NCF) [12] is a reasonable extension of the DeepMF model; NCF replaces the dot layer by a Multi-Layer Perceptron (MLP), providing a deep and non-linear combination of the embedding representations. Both the DeepMF and the NCF models improve the MF results.

RS prediction is a regression task where real values are obtained; however, RS recommendation usually is a classification task, where only a discrete number of fixed values can be returned (e.g. number of stars). Then, deep learning classification approaches naturally fit the CF aims; a classification-based deep learning model [13] is proposed to both implement the recommendation task and provide a reliability value for each recommended item. Additionally, the regular deep classification approach can be improved by combining the obtained <reliability, rating> tuple values [14].

This paper focuses on testing the GANRS [15] generated datasets by applying a representative set of deep learning CF baselines and comparing their recommendation quality results. Generative adversarial networks (GAN) have recently been introduced in the RS area [16] to reinforce the defense strategies of shilling attacks [17], but particularly to improve results by generating augmented data; fake purchase vectors are generated in CFGAN [18] to reinforce the real purchase data. The Wasserstein CFGAN version is the unified GAN (UGAN), and it manages to minimize the GAN collapse mode. Negative sampling information is incorporated in the input data to IPGAN [19], where two different generative models are used, respectively, for positive and negative samples. Temporal patterns have also been combined with GAN models in RecGAN [20], which uses Recurrent Neural Networks (RNN). The reinforcement learning and GAN models are used to process session information rather than rating matrices in the DCFGAN architecture [21]. Conditional rating generation is proposed in [22] by using a Conditional GAN (CGAN). NCGAN [23] uses a GAN to perform recommendation training and a previous neural network stage to obtain the nonlinear features of the users. Finally, unbalanced data sets are processed using the PacGAN concept in the discriminator and a Wasserstein GAN in the generator [24].

Based on Markov chains and recurrent neural networks, RecSim [25] generates synthetic profiles of users and items; its parameterization is low. The social Taobao web site has been used to provide the Virtual-Taobao [26], improving search in this site; internal distributions are simulated by a GAN. RS synthetic data is created using the Java-based generator DataGenCars [27]; it is based on statistical procedures, allowing a flexible parametrization, but returning low accuracy compared to GAN models. Finally, the SynEvaRec [28] framework makes use of the Synthetic Data Vault (SVD) library for RS datasets generation, based on multivariate distributions using copula functions. The SynEvaRec main drawbacks are its poor accuracy and its low performance in the training stage. Table 1 summarizes the existing methods.

|  | method | parameterization | Accuracy | Performance |
|---|---|---|---|---|
| **GANRS** | generative | high | high | high |
| **RecSim** | generative | low | high | high |
| **Virtual-Taobao** | generative | low | middle | high |
| **DataGenCars** | statistical | high | low | high |
| **SynEvaRec** | statistical | high | low | low |

Table I. Comparison table of current RS methods to create CF synthetic data

## A. Main contributions

The objective of this paper is to reinforce the existing tests that have been run on the synthetic datasets generated using the GANRS method. Beyond the existing comparatives between source datasets (Movielens, Netflix, and MyAnimeList) and their synthetic versions, attending to their users, items, and ratings distributions, it is convenient to put into the test the generated datasets on real recommendation scenarios. Some specific and limited prediction and recommendation experiments are provided in the GANRS paper [15], but our research extends them with a comprehensive set of recommendation-based tests, where different deep learning models relevant to the CF are used as baselines and significant recommendation quality measures are processed, and their results are compared.

The paper hypothesis is that the GANRS model can adequately mimic different source CF datasets, such as the Movielens family, MyAnimeList, etc., generating synthetic CF datasets that follow the internal patterns and the probability distributions of the source datasets in the deep learning generative processing. The hypothesis is extended to the different parameterizations the GANRS generative model allows, setting a) the number of fake users, b) the number of fake items, and c) the number of samples. We will put the hypothesis to the test by running different deep learning state of the art CF baselines (NCF, DeepMF, etc.) on several GANRS generated datasets and comparing the obtained recommendation qualities. The GANRS synthetic datasets will contain different number of users, items, and samples. Note that if the hypothesis is fulfilled, the GANRS model can be used as a powerful tool to test current and future CF methods and models on challenging synthetic scenarios where the number of users, items and samples can endlessly grow.

In the rest of the paper, section 2 explains the different deep learning models used in this research, both to generate the synthetic datasets and to test the behavior of baselines on the generated data. Section 3 introduces the experiments design, synthetic datasets, and baselines. It also shows the results obtained, their explanations, and the discussion. Section 4 highlights the main conclusions of the article and the suggested future work. Finally, a references section lists current research in the area.

## 2. Models

This research uses many deep learning models, both the GANRS [15] generative framework with which the synthetic datasets have been obtained and the different models used to test the generated datasets. These baseline models are as follows. DeepMF [10], VDeepMF [11], regression NCF [12], classification NCF [13], improved classification NCF [14] and binary regression.

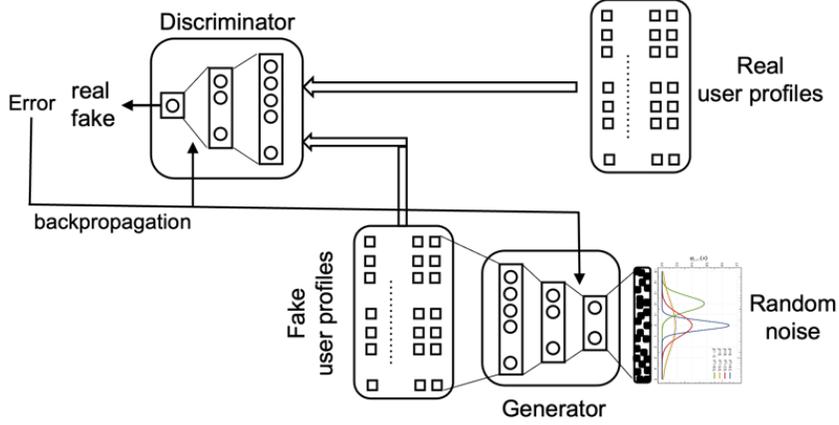

Fig. 1. GANRS architecture.

The GANRS architecture shown in Fig. 1 consists of the generator and discriminator models, where the generator creates CF fake profiles from Gaussian random noise vectors. The discriminator's responsibility is to detect fake samples from training batches of real and fake profiles. Once the RSGAN has been trained from a real source dataset (MovieLens, MyAnimeList, etc.) it can generate as many fake samples as desired by providing the generator with batches of random noise vectors. It is important to note that the GAN is fed with embedded user profiles rather than sparse vectors of ratings. Embeddings are obtained in a previous stage using a DeepMF [10] model.

$$max_D V(D) = E_{x \sim p_{data}(x)}[\log D(x)] + E_{z \sim p_z(z)}\left[\log\left(1 - \{D(G(z))\}\right)\right] \quad (1)$$

where x are real user profiles and z are random noise vectors (Fig. 1).

$$min_G V(G) = \mathbb{E}_{z \sim p_z(z)}[\log(1 - D(G(z)))] \quad (2)$$

$$min_G \, max_D V(D,G) = \mathbb{E}_{x \sim p_{data}(x)}[\log D(x)] + \mathbb{E}_{z \sim p_z(z)}[\log(1 - D(G(z)))] \quad (3)$$

The objective of the discriminator can be defined as its ability to recognize real profiles (first term in (1)) combined with its ability to detect fake profiles (second term in (1)). The generator objective is to generate fake profiles that can fool the discriminator (2). Finally, the GAN can be seen as a minimax game in which the discriminator 'D' tries to maximize V, whereas the generator 'G' tries to minimize it (3).

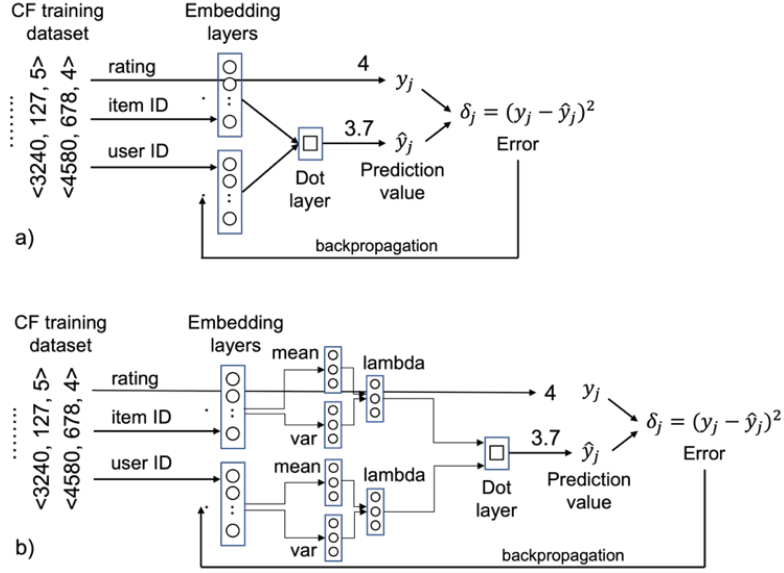

Fig. 2. (a) DeepMF and (b) VDeepMF models.

On an RS dataset containing U users and I items, the prediction of item i to user u is shown in (4), where the function f is defined as a neural network that converts their integer inputs user u ID and item i ID in their corresponding prediction. P and Q denote the neural network equivalence to the hidden factors of the MF, where K is the number of factors (i.e., the number of neurons in each embedding layer). Note that, usually, the set of weights in P and Q are called θ. The prediction of an item i to the user u is computed as the dot product of the embedding layer activations g(u|P) and h(i|Q) in (5). Finally, the squared loss is used (6) to learn the model parameters.

The VDeepMF architecture is an extension of the DeepMF one, where a variational stage is added. Fig. 2b shows the variational stage located between the embedding layers and the dot layer. This variational stage converts input embeddings to parameters of a statistical distribution (usually a Gaussian one). This concept can be seen in the 'mean' and 'variance' layers that follow the VDeepMF embedding layers, both for users and items (Fig. 2b). Each pair of mean and variance layers codes the corresponding Gaussian distribution parameters. Each Lambda layer uses the Gaussian mean and variance to stochastically sample vectors in the latent space. The result is a more robust model due to its stochastic learning.

$$\hat{y}_{ui} = f(P, u, Q, i | P, Q), \text{where } P \in \mathbb{R}^{U \cdot K}, Q \in \mathbb{R}^{K \cdot N} \tag{4}$$

$$\hat{y}_{ui} = dot\big(g(u|P), h(i|Q)\big) \tag{5}$$

$$L_{sqr} = \sum_{(u,i)} (y_{ui} - \hat{y}_{ui})^2 \tag{6}$$

$$(g(u|P), h(i|Q)) = (v_u, w_i) \mapsto \big(\mu_1(v_u), \sigma_1^2(v_u), \mu_2(w_i), \sigma_2^2(w_i)\big) \in \mathbb{R}^{4K} \tag{7}$$

$$\big(P_{\mu_1(v_u), \sigma_1^2(v_u)}, Q_{\mu_2(w_i), \sigma_2^2(w_i)}\big) \tag{8}$$

$$P \sim \mathcal{N}\big(\mu_1(v_u), diag\, \sigma_1^2(v_u)\big), \quad Q \sim \mathcal{N}\big(\mu_2(w_i), diag\, \sigma_2^2(w_i)\big) \tag{9}$$

Equation (7) shows the 'mean' and 'var' layers conversion from embedding latent vectors to activation maps representing Gaussian distributions. Thus, the input of the Lambda layers are the pairs of random vectors in equation (8). In equation (9), $\mathcal{N}$ denotes a K-dimensional multivariate distribution, where μ represents the mean vector and diag σ is the covariance matrix.

Fig. 3. (a) Regression NCF, and (b) Classification NCF.

The Keras code that summarizes each of the baseline models is provided in Table II. Please note that NCF Binary regression can be coded in a similar way to the regular NCF classification, by replacing the size of the deepest layer to only one output neuron.

| **DeepMF** |
|---|
| ```
movie_input = Input(shape=[1],name='movie-input')
movie_embedding = Embedding(num_movies + 1, latent_dim,
      name='movie-embedding')(movie_input)
movie_vec = Flatten(name='movie-flatten')(movie_embedding)
user_input = Input(shape=[1],name='user-input')
user_embedding = Embedding(num_users + 1, latent_dim,
      name='user-embedding')(user_input)
user_vec = Flatten(name='user-flatten')(user_embedding)
dot = Dot(axes=1)([movie_vec, user_vec])
model_DeepMF_regression = Model([user_input, movie_input], dot)
``` |
| **VDeepMF** |
| ```
movie_input = Input(shape=[1],name='movie-input')
movie_embedding = Embedding(num_movies + 1, latent_dim,
        name='movie-embedding')(movie_input)
movie_vec = Flatten(name='movie-flatten')(movie_embedding)
user_input = Input(shape=[1],name='user-input')
user_embedding = Embedding(num_users + 1, latent_dim,
        name='user-embedding')(user_input)
user_vec = Flatten(name='user-flatten')(user_embedding)
concat = Concatenate(axis=1,name='movie-user-concat')
          ([movie_vec, user_vec])
fc_1 = Dense(70, name='fc-1', activation='relu')(concat)
fc_1_dropout = Dropout(0.5, name='fc-1-dropout')(fc_1)
fc_2 = Dense(30, name='fc-2', activation='relu')(fc_1_dropout)
fc_2_dropout = Dropout(0.4, name='fc-2-dropout')(fc_2)
``` |

|   |   |
|---|---|
| | `fc_3 = Dense(1, name='fc-3', activation='relu')(fc_2_dropout)`<br>`model_NCF_regression = Model([user_input, movie_input], fc_3)` |
| **NCF Regression** | |
| | `movie_input = Input(shape=[1],name='movie-input')`<br>`movie_embedding = Embedding(num_movies + 1,`<br>`    latent_dim, name='movie-embedding')(movie_input)`<br>`movie_vec = Flatten(name='movie-flatten')(movie_embedding)`<br>`user_input = Input(shape=[1],name='user-input')`<br>`user_embedding = Embedding(num_users + 1,`<br>`    latent_dim, name='user-embedding')(user_input)`<br>`user_vec = Flatten(name='user-flatten')(user_embedding)`<br>`concat = Concatenate(axis=1,name='movie-user-concat')`<br>`        ([movie_vec, user_vec])`<br>`fc_1 = Dense(50, name='fc-1', activation='relu')(concat)`<br>`fc_1_dropout = Dropout(0.4, name='fc-1-dropout')(fc_1)`<br>`fc_2 = Dense(20, name='fc-2', activation='relu')(fc_1_dropout)`<br>`fc_2_dropout = Dropout(0.4, name='fc-2-dropout')(fc_2)`<br>`fc_3 = Dense(1, name='fc-3',`<br>`    activation='sigmoid')(fc_2_dropout)`<br>`model_regression_relevant = Model([user_input, movie_input], fc_3)` |
| **NCF Classification** | |
| | `movie_input = Input(shape=[1],name='movie-input')`<br>`movie_embedding = Embedding(num_movies + 1, latent_dim,`<br>`    name='movie-embedding')(movie_input)`<br>`movie_vec = Flatten(name='movie-flatten')(movie_embedding)`<br>`user_input = Input(shape=[1],name='user-input')`<br>`user_embedding = Embedding(num_users + 1, latent_dim,`<br>`    name='user-embedding')(user_input)`<br>`user_vec = Flatten(name='user-flatten')(user_embedding)`<br>`concat = Concatenate(axis=1,`<br>`    name='movie-user-concat')([movie_vec, user_vec])`<br>`fc_1 = Dense(80, name='fc-1', activation='relu')(concat)`<br>`fc_1_dropout = Dropout(0.6, name='fc-1-dropout')(fc_1)`<br>`fc_2 = Dense(25, name='fc-2', activation='relu')(fc_1_dropout)`<br>`fc_2_dropout = Dropout(0.4, name='fc-2-dropout')(fc_2)`<br>`fc_3 = Dense(6, name='fc-3', activation='softmax')(fc_2_dropout)`<br>`model_NCF_classification = Model([user_input, movie_input], fc_3)` |

Table II. Keras code of the baseline models

The 'regression NCF' term refers to the regular Neural Collaborative Filtering model. This model extends the DeepMF one by adding a Multi-Layer-Perceptron (MLP) stage, as it can be seen in Fig. 3a. The DeepMF model generates accurate embedding vectors, but it combines them (the user and item vectors) using a linear dot layer. The NCF approach improves the DeepMF model, due to the non-linear and deep learning processing of the embedding output vectors.

$$o_1 = \phi_1(p_u v_u, q_i w_i) \tag{10}$$

$$o_2 = \phi_2(W_2^T o_1 + b_2) \tag{11}$$

$$o_n = \phi_n(W_n^T o_{n-1} + b_{n-1}) \tag{12}$$

$$<r,v> = softmax(W_n^T o_n + b_n) \tag{13}$$

$$\hat{y}_{ui} = v \mid <r,v> \in argmax(r) \tag{14}$$

$$\hat{y}_{ui} = \sum_{r=1}^{R} r \cdot p \tag{15}$$

The additional MLP model is formalized in equations (10) to (12), where $p_u$ and $q_i$ denote the embedding layers weights, $W_x^T$ and $b_x$ are the weight matrix and bias vector of layer x in the MLP, $\phi_x$ denotes the layer x with its activation function. The regression NCF model has an output layer containing a unique neuron with an activation function that is linear, implementing the required regression. In contrast, the NCF classification model replaces this output layer with a layer containing as many neurons as possible votes in the RS (usually from one to five stars), as can be seen in FIG. 3b. The softmax activation function is used in this output layer, while the model loss function is the categorical cross entropy; this ensures a probabilistic output that can be interpreted as a set of <reliability, vote> tuples (13), where the argmax(reliability) selects the predicted vote (14). The improved classification model basically combines the existing <reliability, vote> tuple values (15), providing a more accurate output function than the argmax one.

By combining the GANRS generated datasets with the chosen deep learning baselines and the selected recommendation quality measures, a set of experiments is designed and tested in the next section. Results are shown and explained, and finally an overall discussion is provided.

## 3. Experiments And Results

This paper runs a complete set of experiments to test the performance of current CF deep learning models on GANRS generated datasets.

Table III shows a summary of the designed experiments. The tested CF datasets are generated using 'GANRS' [15], obtained from the source datasets: Netflix* [29], MyAnimeList [30], and Movielens 100K [31]. For comparative reasons, results using the three source datasets are also provided. The six deep learning models chosen as baselines are DeepMF [10] and regression NCF [12], and their variations VDeepMF [11], and classification based NCF [13]. Finally, the 'improved classification NCF' [14] and the binary regression are included. Since we use classification-based models, where recommendations are not a subset of predictions, only recommendation quality measures can be properly used, from which precision, recall, and F1 have been selected. Finally, we have set even values from 2 to 10 as the number of recommendations (N), and the two most relevant rating values as relevancy threshold (θ): 4 & 5 for Movielens and Netflix*, and 9 & 10 for MyAnimeList.

| CF deep learning models | CF Datasets | Quality Measures | Testing parameters |
|---|---|---|---|
| DeepMF [10] | Netflix* [29] | Precision | Relevance threshold (θ): |
| VDeepMF [11] | GANRS Netflix*: 2,000; 8,000 users | Recall | 9, 10 (MyAnimeList): |
| | | | 4, 5 (Netflix* and Movielens). |
| Regression NCF [12] | GANRS Netflix*: 150K, 500K, 3M | F1 | |
| | | | Number of recommendations (N): |
| Classification NCF [13] | Movielens 100K [31] | | [2, 4, 6, 8, 10] |
| Classification improved NCF [14] | GANRS Movielens 100K: 2,000; 8,000 users | | |
| | | | Gaussian standard deviation: 2.5 |
| Binary regression | MyAnimeList [30] | | |
| | GANRS MyAnimeList: 2,000; 8,000 users | | |

Table III. Information Summary Of The Designed Experiments

| Dataset | #users | #items | #ratings | scores | sparsity |
|---|---|---|---|---|---|
| Movielens 100K | 943 | 1682 | 99,831 | 1 to 5 | 93.71 |
| Netflix* | 23,012 | 1,750 | 535,421 | 1 to 5 | 98.68 |
| MyAnime | 19,179 | 2,692 | 548,967 | 1 to 10 | 98.94 |
| GANRS Netflix* 2,000 | 2,000 | 4,000 | 405,539 | 1 to 5 | 94.93 |
| GANRS Netflix* 8,000 | 8,000 | 4,000 | 628,194 | 1 to 5 | 98,03 |
| GANRS Netflix* 150K | 2,000 | 4,000 | 108,710 | 1 to 5 | 98,64 |
| GANRS Netflix* 500K | 2,000 | 4,000 | 272,853 | 1 to 5 | 96,59 |
| GANRS Netflix* 3M | 2,000 | 4,000 | 587,651 | 1 to 5 | 92,65 |
| GANRS Movielens 2,000 | 2,000 | 4,000 | 353,269 | 1 to 5 | 95,58 |
| GANRS Movielens 8,000 | 8,000 | 4,000 | 509,193 | 1 to 5 | 98,40 |
| GANRS MyAnime 2,000 | 2,000 | 4,000 | 419,234 | 1 to 10 | 94,76 |
| GANRS MyAnime 8,000 | 8,000 | 4,000 | 654,247 | 1 to 10 | 97,95 |

Table IV. Main Parameter Values Of The Tested Datasets

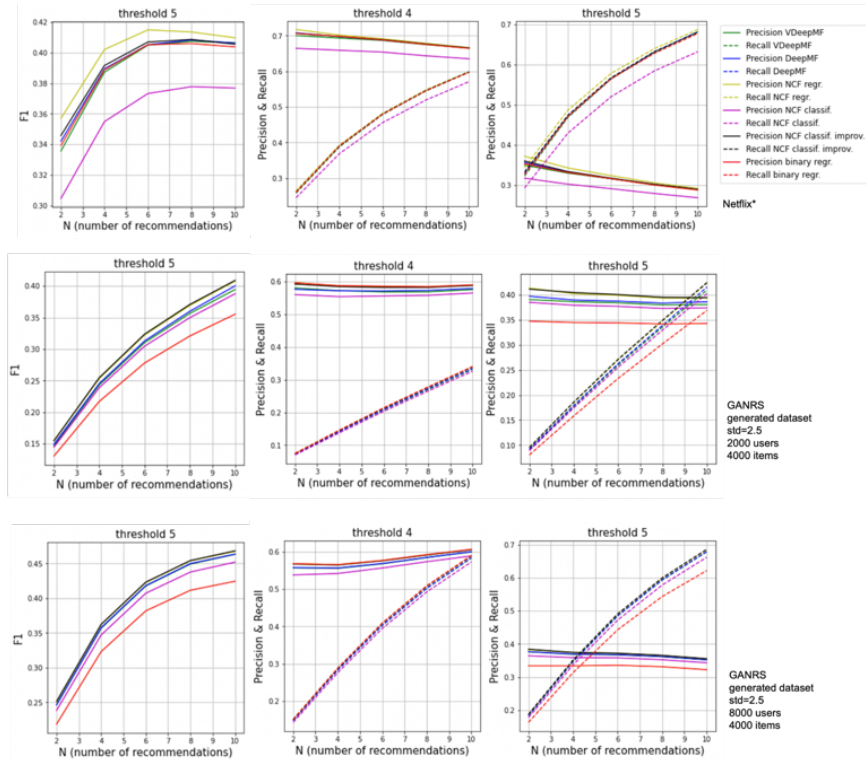

Fig. 4. Comparative among Netflix*, GANRS 2,000 users, and GANRS 8,000 users. Generated datasets include 4000 items and sets 2.5 for the standard deviation of the Gaussian random noise. Number of recommendations N = [2, 4, 6, 8, 10].

Table IV shows the values of the main parameters for both the real and synthetic datasets used in the designed experiments. Our first set of experiments are based on the source dataset Netflix*, and it compares the quality recommendation results obtained both from Netflix* and their synthetic generated versions: 2,000 & 8,000 users.

The three rows in Fig. 4 show, respectively, the results on Netflix* (top row), on GANRS 2,000 users (middle row), and on GANRS 8,000 users (bottom row). The middle and right columns show the precision and recall values when threshold θ is set to 4 and 5 (respectively). The left column shows the precision/recall based F1 quality measure. The legend in the upper-right area of Fig. 4 holds the colors that represent each one of the chosen deep learning baselines. Note that the expected behavior is the superior performance of the deep learning models: regression NCF, improved NCF classification, VDeepMF and DeepMF, whereas classification NCF and binary regression should provide weaker results.

### A. Experiment 1: Netflix* versus GANRS 2000 users, versus GANRS 8000 users

This experiment compares the absolute values and the trends in the recommendation quality obtained for each baseline when applied to the original Netflix* dataset, to the GANRS generated dataset setting 2000 users, and to the GANRS generated dataset setting 8000 users. Both generated datasets take Netflix* as the source to catch its internal patterns. We expect similar trends in the graph functions, showing that the GANRS generated datasets adequately mimic the Netflix* patterns. We also expect different absolute quality values due to the different number of users selected for each GANRS generated dataset.

The top row in Fig. 4 (Netflix*) shows the expected performance evolutions, where the higher the number of recommendations (x-axis), the lower the prediction quality measure, and the higher the recall (it is more complicated to get right 10 recommendations than to get right the two most promising ones). In the same way, a lower threshold value (middle graph) gets a better precision than a higher threshold value (right graph), since there are more samples that reach the threshold, and consequently it is easier to get right with the recommended items. In contrast, the higher the threshold, the better the recall, since there will be less relevant items in the recall denominator. Once we have checked the expected behaviors, the key question is: will the synthetic datasets accomplish the expected trends? Looking at the middle and bottom rows in Fig. 4 we can observe the same aforementioned tendency. The relevant difference between the results from the source Netflix* and the generated GANRS is not the quality trend, but the absolute precision and recall values, where the precision is slightly superior in the GANRS datasets, whereas recall is lower. Please note that the Netflix* dataset contains 23,012 users (Table IV), and then the GANRS versions, particularly the 2000 user version, suffer from a lack of richness that influences the recall results. Additionally, as expected, the higher the threshold, the worse the precision and the better the recall.

The F1 quality measure (left column in Fig. 4) balances precision and recall and allows us to visually compare the different results of the data set. We can observe that synthetic datasets provide quality trends and values that are compatible with those achieved by their source dataset (Netflix*). In addition, the GANRS 8,000 user results are more similar to the source than the GANRS 2,000 user ones, as expected due to the 23,000 users contained in Netflix*. Regarding the behavior of deep learning baselines, synthetic data sets maintain the 'ranking order' obtained from Netflix*, where the regression NCF slightly 'wins', closely followed by the improved classification NCF, DeepMF and VDeepMF. NCF classification and binary regression swap their position in the queue when tested on Netflix* and GANRS. Overall, synthetic GANRS datasets perform adequately for CF testing using state-of-the-art deep learning models.

### B. Experiment 2: Netflix* based GANRS 3 million samples, versus GANRS 500 thousand samples, versus GANRS 150 thousand samples

This experiment compares the absolute values and the trends in the recommendation quality obtained for each baseline when applied to the GANRS generated dataset setting 3 million samples, to the GANRS generated dataset setting 500 thousand samples, and to the GANRS generated dataset setting 150 thousand samples. All the generated datasets take Netflix* as the source to catch its internal patterns. Like the previous experiment, we expect similar trends in the graph functions, showing that the GANRS generated datasets adequately mimic the Netflix* patterns. We also expect different absolute quality values due to the different number of samples selected for each GANRS generated dataset.

The following experiment uses three synthetic GANRS datasets where the number of samples varies. We use the GANRS Netflix* 150K, 500K, and 3M versions (Table II) which, respectively, contain 108710, 272853 and 587651 samples. Fig. 5 shows the recommendation results obtained in the 3M version (top row), the 500K version (middle row), and the 150K version (bottom row). As expected, precision decreases as size falls; this effect can be particularly observed in the most extreme experiment: the highest threshold (θ =5) combined with the smallest dataset (150K version). On the other hand, the larger the dataset, the lower the recall results, since there will be more 'total relevant' items in each recommendation process. This effect is more severe when the threshold is not high (θ =4), since even further 'total relevant' items will be in the denominator of the recall quality measure. The top and middle graphs of the 'threshold 4' column in Fig. 5 show the concept. Beyond the specific quality values, we can observe that it is possible to use generated datasets with different sizes to test CF machine learning models in different scenarios: the results will show the expected behavior and trends. Regarding the tested deep learning models, it is interesting to observe how the NCF classification and, particularly, the binary regression dramatically decreases their performance when the dataset size increases. We can also see how the improved NCF classification reaches the NCF regression, compared to the results in Fig. 4.

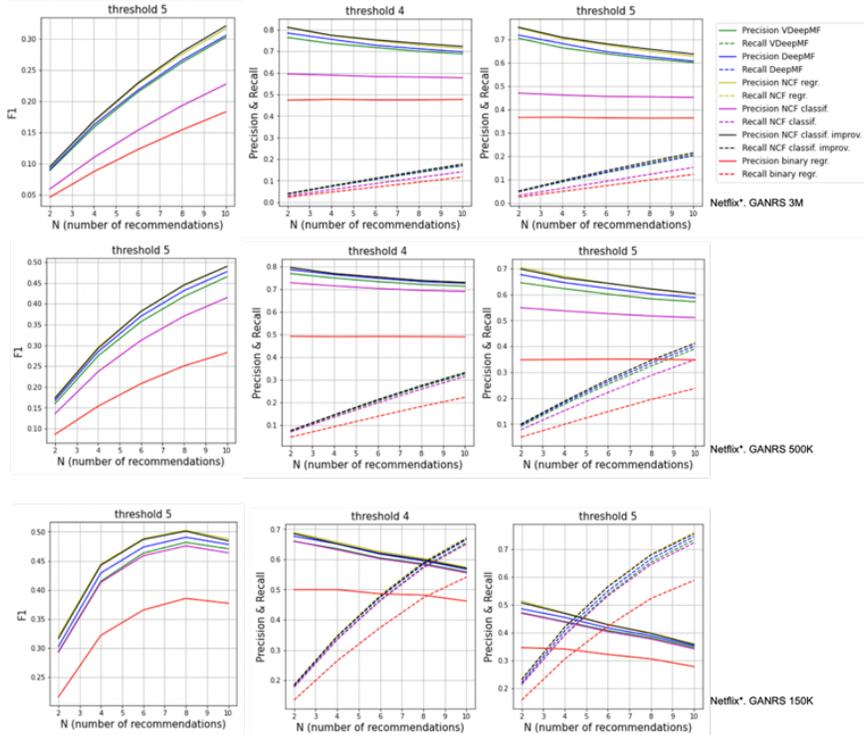

Fig. 5. Comparative among Netflix*, GANRS 150K, 500K, and 3M samples. Generated datasets with 2,000 users; 4,000 items and 2.5 for the standard deviation of the Gaussian random noise. Number of recommendations N = [2, 4, 6, 8, 10].

## C. Experiment 3: Movielens 100K versus GANRS 2000 users, versus GANRS 8000 users

This experiment compares the absolute values and the trends in the recommendation quality obtained for each baseline when applied to the source Movielens 100K dataset, to the GANRS generated dataset setting 2000 users, and to the GANRS generated dataset setting 8000 users. Both generated datasets take Movielens 100K as the source to catch its internal patterns. As in the previous subsections, we expect similar trends in the graph functions, showing that the GANRS generated datasets adequately mimic the Movielens 100K patterns. We also expect different absolute quality values due to the different number of users selected for each GANRS generated dataset.

To avoid unnecessary repetitions, experiments on the synthetic datasets generated from Movielens and MyAnimeList are restricted to the 2,000 versus 8,000 user comparatives.

Fig. 6 shows the Movielens results; they are similar to those obtained using generated datasets from Netflix*. In fact, both sets of synthetic data contain a similar number of samples: 405,539 versus 353,269 in the 2,000 user versions and 628,194 versus 509,193 in the 8,000 user datasets. Comparing the precision & recall results of the GANRS versions, both at thresholds 4 and 5 in Fig. 4 and 6, we can see that the absolute values (y-axis) and the curve trends are similar. Regarding the baselines, the NCF regression provides a balanced (F1) superiority, as it happens in the source Netflix* data set.

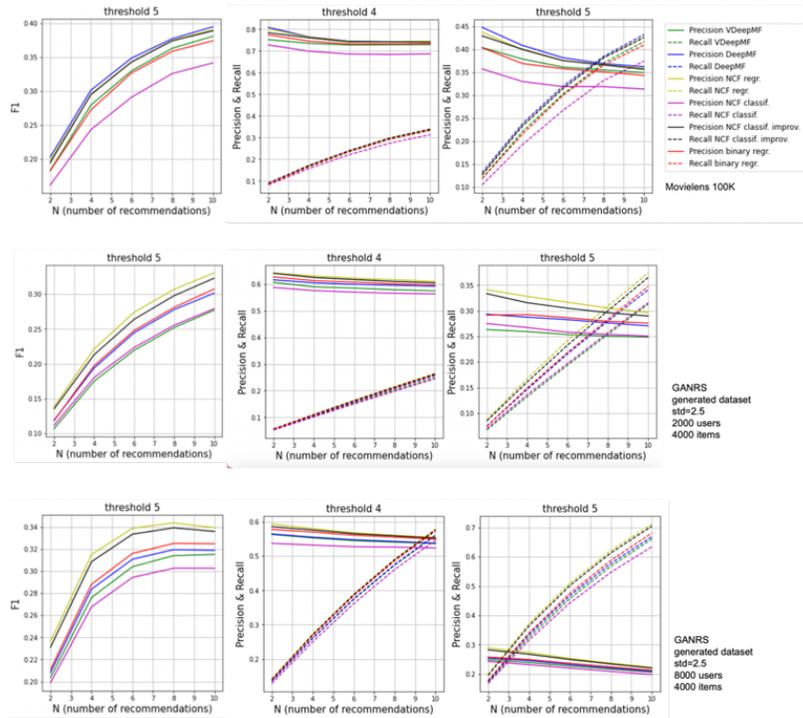

Fig. 6. Comparative among Movielens 100K, GANRS 2,000 users; and GANRS 8,000 users. Generated datasets with 4,000 items and 2.5 for the standard deviation of the Gaussian random noise. Number of recommendations N = [2, 4, 6, 8, 10].

## D. Experiment 4: MyAnimeList versus GANRS 2000 users, versus GANRS 8000 users

This experiment compares the absolute values and the trends in the recommendation quality obtained for each baseline when applied to the MyAnimeList dataset, to the GANRS generated dataset setting 2000 users, and to the GANRS generated dataset setting 8000 users.

The MyAnimeList family of generated datasets provides interesting results, since MyAnimeList contains a range of ten ratings (1 to 10) instead of the usual 1 to 5. Focusing on the threshold ($\theta=10$) in Fig. 7, it can be observed that precision improves (compared to the preceding results when $\theta=5$). It probably happens due to a higher proportion of ratings 10, compared to the equivalent (ratings 5) in Movielens or Netflix*. The important here is that the synthetic datasets in Fig. 7 mimic this behavior; that is: the comparative between MyAnimeList (Fig. 7 top right graph) and Movielens/Netflix* (Fig. 4 and 6 top-right graphs), looks similar to the comparative between the MyAnimeList GANRSs (Fig. 7 middle-right and bottom-right graphs) and Movielens/Netflix* GANRSs (Fig. 4 and 6 middle-right and bottom-right graphs). This means that the GANRS synthetic datasets are adequate. Finally, as expected, the classification models perform worst in this scenario (exception the improved one), since it is harder to correctly classify ten categories than five categories.

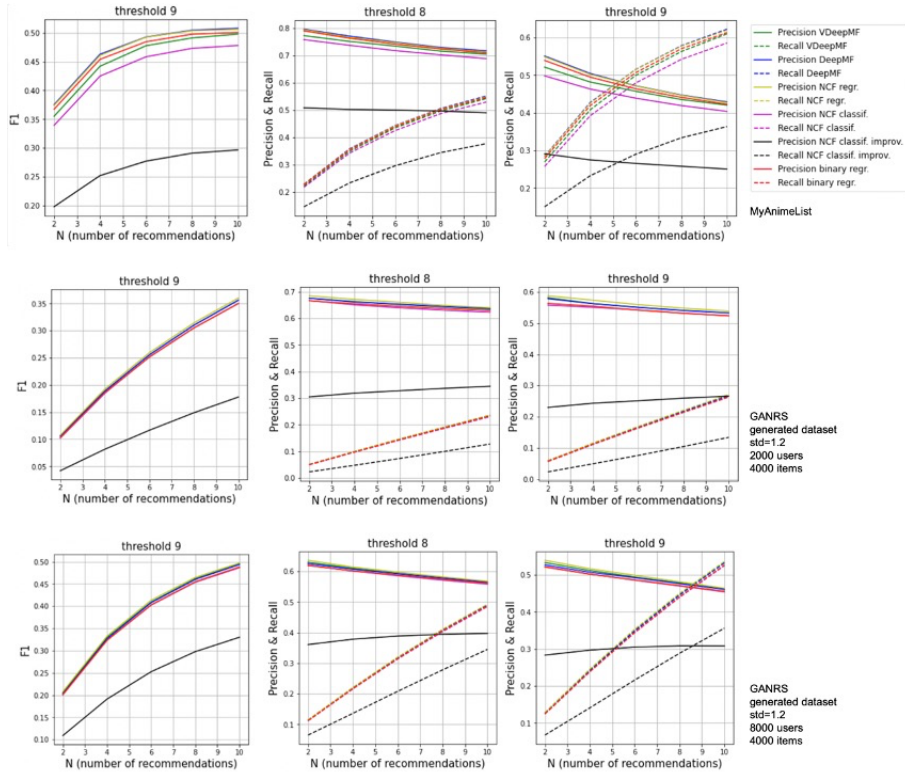

Fig. 7. Comparative among MyAnimeList, GANRS 2,000 users; and GANRS 8,000 users. Generated datasets with 4,000 items and 2.5 for the standard deviation of the Gaussian random noise. Number of recommendations N = [2, 4, 6, 8, 10].

## 4. Discussion

Overall, the obtained results show that the synthetic GANRS datasets adequately mimic the behavior of the source datasets from which the GAN learns their patterns. Results sustain the hypothesis of the paper, and they confirm that the GANRS generator creates synthetic datasets containing similar patterns and probability distributions to the chosen source datasets, and what is more: this is also true when the selected number of users, items and samples varies. Our view is that the GANRS generative model gets its successful behavior from the architectural key with which it has been designed: to feed the GAN kernel of the model with short and dense embeddings instead of the traditional large and sparse raw data [15]. In this way the GAN stage improves its performance, better catches the source patterns, and it reduces the mode collapse condition.

Since the synthetic datasets can be generated setting their sizes, number of items, and number of users, it is possible to use them to test CF machine learning models on different scenarios, e.g., when the number of users varies. Specifically, all the synthetic datasets tested in the experiments show adequate variation of precision and recall, where precision improves (and recall gets worst) as the number of samples increases. This is because the higher the total number of samples, the higher the average number of ratings for each user. Additionally, as expected, accuracy and recall differ when tested by setting different recommendation thresholds. Observing the results of the tested deep learning models, the NCF regression and the improved NCF classification perform significantly better than the NCF classification and the binary regression. DeepMF and VDeepMF provide slightly lower quality results than NCF regression. All these results are compatible with the state-of-the-art ones. Finally, it is remarkable how the tested GANRS datasets adequately catch the quality loss of the NCF classification when MyAnimeList is taken as a source, since this dataset encodes a ten ratings interval instead of the usual five ratings interval.

## 5. Conclusions

This paper tests the performance of the synthetic datasets generated from the published GANRS method. A representative set of generated datasets has been created by selecting a different number of users and a different number of samples. The obtained datasets have been tested on six representative CF deep learning models:

DeepMF, VDeepMF, NCF, NCF classification, improved NCF classification, and binary regression. The recommendation quality measures precision, recall, and F1 have been chosen. The results show adequate performance of the synthetic datasets on all applied deep learning models. In particular, it can be observed that, as expected, precision improves when the size of the dataset increases, as well as when the average number of ratings of each user also increases. In the same way, the recall decreases as the size of the data set increases. The interval of the ratings in the dataset (ten in MyAnimeList and five in Movielens and Netflix*) has the expected impact, where both the recall and, particularly, the precision drop using MyAnimelist. The tested CF deep learning models perform similarly when the results of the synthetic datasets are compared with the real datasets, and it happens on the different combinations of the selected number of samples and number of users. Overall, the GANRS method generates valuable synthetic datasets that can be used to test new deep learning models proposed in the CF RS area. Future works include testing synthetic datasets tailored to specific CF scenarios such as user cold start, item cold start, dataset cold start, imbalanced data, demographic variations, binary ratings (like, non-like), fairness, recommendation to groups of users, and heavy sparse data.

## Acknowledgements

This work was partially supported by *Ministerio de Ciencia e Innovación* of Spain under the project PID2019-106493RB-I00 (DL-CEMG) and the *Comunidad de Madrid* under *Convenio Plurianual* with the Universidad Politécnica de Madrid in the actuation line of *Programa de Excelencia para el Profesorado Universitario*.

## Conflict of interest

The authors declare that they have no conflict of interest.

## References

[1] J. Bobadilla, A. González-Prieto, F. Ortega, R. Lara-Cabrera. Deep learning feature selection to unhide demographic recommender systems factors. In: Neural Computing and Applications 2021, vol. 33(12), pp. 7291-7308.

[2] Y. Deldjoo, M. Schedl, P. Cremonesi, G. Pasi. Recommender systems leveraging multimedia content. In: ACM Computing Surveys (CSUR), 2020, vol. 53(5), pp. 1-38.

[3] S. Kulkarni, S.F. Rodd. Context aware recommendation systems: A review of the state of the art techniques. In: Computer Science Review 2020, vol.37, 100255.

[4] J. Shokeen, C. Rana. A study on features of social recommender systems. In: Artificial Intelligence Review 2020, vol. 53(2), pp. 965-988.

[5] J.B. Schafer, D. Frankowski, J. Herlocker, S. Sen. Collaborative Filtering Recommender Systems. In: The Adaptive Web. Lecture Notes in Computer Science, Brusilovsky, P., Kobsa, A., Nejdl, W. (eds), Springer, Berlin, Heidelberg, 2007, vol. 4321.

[6] E. Cano, M. Morisio. Hybrid recommender systems: A systematic literature review. In: Intelligent Data Analysis, 2017, vol. 21(6), pp. 1487-1524.

[7] B. Zhu, R. Hurtado, J. Bobadilla, F. Ortega. An efficient recommender system method based on the numerical relevances and the non-numerical structures of the ratings. In: IEEE Access, 2018, vol. 6, pp. 49935-49954.

[8] Mnih, R. R. Salakhutdinov, Probabilistic matrix factorization. In: Advances in neural information processing systems, 2007, vol. 20.

[9] F'evotte, J. Idier. Algorithms for nonnegative matrix factorization with the Œ≤-divergence. Neural computation, 2011, vol. 23(9), pp. 2421-2456.

[10] Hong-Jian Xue, Xinyu Dai, Jianbing, Zhang, Shujian Huang, Jiajun Chen. Deep Matrix Factorization Models for Recommender Systems. In: Proceedings of the Twenty-Sixth International Joint Conference on Artificial Intelligence, 2017, pp. 3203-3209.

[11] J. Bobadilla, J. Dueñas, A. Gutiérrez, F. Ortega. Deep Variational Embedding Representation on Neural Collaborative Filtering Recommender Systems. In: Applied Sciences, 2022, vol. 12(9), 4168.

[12] Xiangnan He, Lizi Liao, Hanwang Zhang. Neural Collaborative Filtering. In: International World Wide Web Conference Committee (IW3C2), 2017, pp. 173-182.

[13] J. Bobadilla, F. Ortega, A. Gutiérrez, S. Alonso. Classification-based Deep Neural Network Architecture for Collaborative Filtering Recommender Systems. In: International Journal of Interactive Multimedia and Artificial Intelligence, 2020, vol. 6(1), pp. 68-77.

[14] J. Bobadilla, A. Gutiérrez, S. Alonso, A. González-Prieto. Neural Collaborative Filtering Classification

Model to Obtain Prediction Reliabilities. In: International Journal of Interactive Multimedia and Artificial Intelligence, 2022, vol. 7(4), pp. 18-26.

[15] J. Bobadilla, A. Gutiérrez, R. Yera, L. Martínez. Creating Synthetic Datasets for Collaborative Filtering Recommender Systems using Generative Adversarial Networks, 2023, arXiv:2023.01297v1

[16] Min Gao, Junwei Zhang, Junliang Yu, Jundong Li, Junhao Wen, Qingyu Xiong. Recommender systems based on generative adversarial networks: A problem-driven perspective. In: Information Sciences, 2021, vol. 546, pp. 1166-118.

[17] Y. Deldjoo; T. Noi, F.A. Merra. A Survey on Adversarial Recommender Systems: From Attack/Defense Strategies to Generative Adversarial Networks. In: ACM computing surveys, 2021, vol. 54 (2), pp. 1-38.

[18] Chae, J. Kang, S. Kim, J. Lee. CFGAN: a generic collaborative filtering framework based on generative adversarial networks. In: Proceedings of the 27th, ACM International Conference on Information and Knowledge Management, CIKM 2018, 2018, pp. 137-146.

[19] G. Guo, H. Zhou, B. Chen, et al. IPGAN: Generating informative item pairs by adversarial sampling. In: IEEE Transactions on Neural Networks and Learning Systems, 2022, vol. 33(2), pp. 694-706.

[20] H. Bharadhwaj, H. Park, B.Y. Lim. Recgan: recurrent generative adversarial networks for recommendation systems. In: Proceedings of the 12th ACM, Conference on Recommender Systems, RecSys 2018, 2018, pp. 372-376.

[21] J. Zhao, H. Li, L. Qu, Q. Zhang, Q. Sun, H. Huo, M. Gong. DCFGAN: An adversarial deep reinforcement learning framework with improved negative sampling for session-based recommender systems. In: Information sciences, 2022, vol. 596, pp. 222-235.

[22] J. Wen, X. Zhu, C.D. Wang, Z. Tian. A framework for personalized recommendation with conditional generative adversarial networks. In: Knowledge and information systems, 2022, vol. 64(10), pp. 2637-2660.

[23] J. Sun, B. Liu, H. Ren, W. Huang. NCGAN: A neural adversarial collaborative filtering for recommender system. In: Journal of intelligent & fuzzy systems, 2022, vol. 42(4), pp. 2915-2923.

[24] W. Shafqat, Y.C. Byun. A Hybrid GAN-Based Approach to Solve Imbalanced Data Problem in Recommendation Systems. In: IEEE access, 2022, vol. 10, pp. 11036-11047.

[25] M. Mladenov, C.W. Hsu, V. Jain, E. Ie, C. Colby, N. Mayoraz, H. Pham, D. Tran, I. Vendrov, C. Boutilier. Demonstrating Principled Uncertainty Modeling for Recommender Ecosystems with RecSim NG. In: RecSys 2020 - 14th ACM Conference on Recommender Systems, pp. 591–593.

[26] J.C Shi, Y. Yu, Q. Da, S.Y. Chen, A.X. Zeng. Virtual-Taobao: Virtualizing real-world online retail environment for reinforcement learning. In: 33rd AAAI Conference on Artificial Intelligence, AAAI 2019, 31st Innovative Applications of Artificial Intelligence Conference, IAAI 2019 and the 9th AAAI Symposium on Educational Advances in Artificial Intelligence, EAAI 2019, pp. 4902–4909. arXiv:1805.10000.

[27] M. del Carmen, S. Ilarri, R. Hermos, R. Trillo-Lado, R. Datagencars: A generator of synthetic data for the evaluation of contextaware recommendation systems. In: Pervasive and Mobile Computing, 2017, vol. 38, pp. 516–541.

[28] V. Provalov, E. Stavinova and P. Chunaev. SynEvaRec: A Framework for Evaluating Recommender Systems on Synthetic Data Classes. In: International Conference on Data Mining Workshops (ICDMW), Auckland, New Zealand, 2021, pp. 55-64.

[29] F. Ortega, B. Zhu, J. Bobadilla, A. Hernando. CF4J: Collaborative filtering for Java. In: Knowledge-Based Systems, 2018, vol. 152, pp. 94-99.

[30] https://www.kaggle.com/azathoth42/myanimelist

[31] F.M. Harper, J.A. Konstan. The movielens datasets: History and context. In: ACM Transactions on Interactive Intelligent Systems, 2015, vol. 5(4), pp. 1-19.